\documentstyle[12pt,aasms4]{article}
\lefthead{Meyer, Cardelli, \& Sofia}
\righthead{Abundance of Interstellar Nitrogen}

\begin{document}

\title{The Abundance of Interstellar Nitrogen\altaffilmark{1}}

\author{David M. Meyer}
\affil{Department of Physics and Astronomy, Northwestern University,
Evanston, IL  60208}
\authoremail{meyer@elvis.astro.nwu.edu}

\and

\author{Jason A. Cardelli\altaffilmark{2} and Ulysses J. Sofia}
\affil{Department of Astronomy and Astrophysics, Villanova University,
Villanova, PA  19085}
\authoremail{usofia@ast.vill.edu}

\altaffiltext{1}{Based on observations obtained with the NASA/ESA
{\it Hubble Space Telescope} through the Space Telescope Science
Institute, which is operated by the Association of Universities for
Research in Astronomy, Inc., under NASA contract NASA-26555.}
\altaffiltext{2}{Deceased 1996 May 14.}

\begin{abstract}
Using the {\it HST} Goddard High Resolution Spectrograph (GHRS),
we have obtained high S/N echelle observations of the
weak interstellar N I $\lambda\lambda$1160,1161 absorption
doublet toward the stars $\gamma$ Cas, $\lambda$ Ori, $\iota$ Ori,
$\kappa$ Ori, $\delta$ Sco and $\kappa$ Sco.  In combination
with a previous GHRS measurement of N I toward $\zeta$ Oph,
these new observations yield a mean interstellar gas-phase nitrogen
abundance (per 10$^6$ H atoms) of 10$^6$ N/H $=$ 75 $\pm$ 4
($\pm$1$\sigma$).
There are no statistically significant variations in the measured N
abundances from sightline to sightline and no evidence of
density-dependent nitrogen depletion from the gas phase.  Since N is
not expected to be depleted much into dust grains in these diffuse
sightlines, its gas-phase abundance should reflect the {\it total}
interstellar abundance.  Consequently, the GHRS observations imply
that the abundance of interstellar nitrogen (gas plus grains)
in the local Milky Way is about 80\% of the solar system value of
10$^6$ N/H $=$ 93 $\pm$ 16.  Although this interstellar abundance
deficit is somewhat less than that recently found for oxygen and
krypton with GHRS, the solar N abundance and the N I oscillator
strengths are too uncertain to definitively rule out either a solar
ISM N abundance or a 2/3 solar ISM N abundance similar to that of
O and Kr.
\end{abstract}


\section{Introduction}

Accurate measurements of the elemental abundances in the interstellar
medium are crucial to studies ranging from the chemical evolution
of the Galaxy (\cite{tim95}) to the composition of 
interstellar dust grains (\cite{sno96}).  Since it is difficult to
obtain such data, the traditional approach has been to adopt the solar
system values as ``cosmic'' current-epoch abundance standards.
Recently, sensitive UV measurements of very weak interstellar
absorption lines with the Goddard High Resolution Spectrograph (GHRS)
onboard the {\it Hubble Space Telescope} ({\it HST}) have begun to
seriously challenge these solar standards.  In particular, based on
GHRS observations of the O I $\lambda$1356 absorption in 13 sightlines,
\cite{mey98} have measured a {\it total} (gas plus dust)
abundance of interstellar oxygen that is 2/3 of the solar value.
\cite{car97} have found similar results for interstellar krypton
which is important since Kr, as a noble gas, should not be depleted
much into dust grains.  These findings are also consistent with
the subsolar CNO abundances that have been measured in nearby B stars
and are likely reflective of the current ISM abundance pattern
(\cite{gie92,kil92,cun94,kil94}).  Since the solar system abundances
are presumably representative of the ISM at the time of the Sun's
formation 4.6 Gyr ago and the ISM abundances should slowly increase
over time (\cite{aud76}, Timmes et al.\ 1995), a 2/3 solar standard
for the local ISM today is difficult to understand in
the context of Galactic chemical evolution models.

Among the abundant CNO elements, nitrogen can potentially provide
the best test of a subsolar ISM abundance pattern since it is the
least likely to be significantly depleted into dust grains.
For example, using the wavenumber-integrated cross-section of the
2.96 $\mu$m N--H stretch (\cite{tie91}), the ISO spectrum of the star
VI Cyg No. 12 (\cite{whi97}) limits the solid-state N abundance
in the N--H stretch to 10$^6$ N/H $<$ 1 in
this heavily reddened sightline.  {\it Copernicus}
observations (\cite{fer81,yor83,kee85})
of various N I transitions in the ultraviolet have
yielded a mean interstellar gas-phase N abundance that is 50\%
to 80\% of the solar value (10$^6$ N/H $=$ 93 $\pm$ 16)
(\cite{gre93}).  However,
the scatter in these data is too great to discriminate the 0.2 dex
difference between a solar and a B-star nitrogen abundance.
Since this scatter is at least partially due to the errors in
measuring the weakest and most optically thin N I lines, the greater
sensitivity of GHRS makes it possible to establish a more accurate
set of interstellar nitrogen abundances.  In this {\it Letter},
we present the results of such an effort involving new GHRS
observations of the very weak N I intersystem doublet at 1159.817
and 1160.937 \AA\ toward six stars.

\section{Observations}

Observations of the interstellar N I $\lambda\lambda$1160,1161
absorption toward the stars $\gamma$ Cas, $\lambda$ Ori, $\iota$ Ori,
$\kappa$ Ori, $\delta$ Sco, and $\kappa$ Sco were obtained with
GHRS in 1996 August and 1997 January using the echelle-A grating
and the 2\farcs0 large science aperture.  The observations of each
star consist of multiple FP-Split exposures that are divided into 
four subexposures
taken at slightly different grating positions so as to minimize
the impact of the GHRS Digicon detector's fixed pattern noise (FPN)
on the reduced data.  Each subexposure
was sampled twice per diode at a velocity resolution of
3.5 km s$^{-1}$.

The data were reduced using the \cite{car94a} recipe to maximize
the S/N ratio of GHRS spectra.  In brief, this process involves:
(1) merging the FP-Split subexposures in diode space so as to
create a template of the FPN spectrum, (2) dividing each
subexposure by this FPN spectrum, (3) aligning the rectified
subexposures in wavelength space using the interstellar lines as
a guide, and (4) summing the aligned subexposures to produce the
net N I spectrum of each star.  As illustrated in Figure 1,
the resulting continuum-flattened spectra reveal convincing
detections of the interstellar N I $\lambda$1160 absorption in
all of the six sightlines comprising our sample.  The S/N ratios
of these spectra range from 150 to 250.  Our measured
equivalent widths for the N I $\lambda$1160 and $\lambda$1161
lines are listed in Table 1 along with the previously reported
GHRS measurements toward $\zeta$ Oph (\cite{sav92}).

The N I column densities given in Table 1 were calculated using
the \cite{hib85} oscillator strengths.
The uncertainties in these theoretically-determined
$f$-values should be no more than the quoted 20\%
(Hibbert et al. 1985) since \cite{sof94} have
empirically verified that they are consistent with the accurate
$f$-values appropriate for the stronger N I $\lambda$1200
transitions.  
The $\lambda\lambda$1160,1161
absorption is generally weak enough for $N$(N I) to be confidently
derived under the assumption that the lines are optically thin.
However, based on the relative N I line strengths toward
$\lambda$ Ori and $\delta$ Sco, a slight correction for
saturation was applied using a Gaussian curve-of-growth with
respective $b$-values of 5.0$^{+\infty}_{-2.5}$ and
10.0$^{+\infty}_{-5.0}$ km s$^{-1}$.  The resultant N I column
densities are 5\% and 6\% greater than their weak line limits,
respectively.  
The N I column density uncertainties given in Table 1 reflect
the estimated errors in the measured equivalent widths and the
saturation corrections (where applied).

\section{Discussion}

With an ionization potential of 14.534 eV, N I should be the
dominant ion of N in H I regions, and little N I should
originate from H II regions.  Consequently, the ratio of
$N$(N I) to the total H column density
[$N$(H) $=$ 2$N$(H$_2$) $+$ $N$(H I)] should accurately reflect
the interstellar gas-phase N/H abundance ratio.  The values of
$N$(H) listed in Table 1 were calculated from the H$_2$ column
densities measured by Savage et al.\ (1977) (and Jenkins, Savage,
\& Spitzer 1986 for $\kappa$ Sco) and the weighted means of
the Bohlin, Savage, \& Drake (1978) (Jenkins et al.\ 1986 in the
case of $\kappa$ Sco) and Diplas \& Savage (1994) $N$(H I) data.
The resulting $N$(N I)/$N$(H) ratios for the 7 GHRS sightlines
yield a weighted mean (\cite{bev69})
interstellar gas-phase N abundance of
10$^6$ N/H $=$ 75 $\pm$ 4 ($\pm$1$\sigma$)
that is about 80\% of the Grevesse \&
Noels (1993) solar abundance (10$^6$ N/H $=$ 93 $\pm$ 16).
The spread in the GHRS nitrogen abundances is about $\pm$0.1 dex
with the most discrepant values being those of $\delta$ Sco
and $\kappa$ Sco at 1.6$\sigma$ above and 1.1$\sigma$ below
the mean, respectively.  It is worth noting that these two
sightlines also have the most discrepant $N$(H I) measurements
in our sample.

In the top panel of Figure 2, the interstellar gas-phase N
abundances are plotted as a function of the fractional abundance
of molecular hydrogen, $f$(H$_2$) = $2N$(H$_2$)/$N$(H), in the GHRS 
sightlines.
As discussed by \cite{car94b}, this parameter separates
sightlines rather distinctly into groups with low and high
$f$(H$_2$) values that are indicative of the physical differences
between UV transparent and H$_2$ self-shielding environments.
Since the former type of environment is typically less hospitable
to grains, higher gas-phase abundances of an element in the low
$f$(H$_2$) group than in the high group is a sign of both the
presence of that element in dust and changes in the elemental
dust abundance due to grain growth and/or destruction.  Figure 2
clearly shows that the gas-phase abundance of interstellar N
does not increase with decreasing $f$(H$_2$) and is thus
consistent with the expectation that nitrogen is not depleted
much into dust grains.

In the bottom panel of Figure 2, the interstellar gas-phase
N/Kr abundance ratio is plotted as a function of $f$(H$_2$)
for the four GHRS sightlines in common between this study and
that of \cite{car97}.  Krypton can be used as a hydrogen-like
benchmark in interstellar abundance studies since it should
not be depleted into grains and Kr/H exhibits a tight spread
of $\pm$0.05 dex among the ten sightlines studied by \cite{car97}.
Although the N/Kr sample is too small for definitive conclusions,
it does appear from Figure 2 that the spread in N/Kr is
tighter than that in N/H\@.  In particular, the sightline
($\delta$ Sco) that stands out the most with a solar abundance in
terms of N/H drops back to the pack in terms of N/Kr\@.  The
most likely explanation for this behavior is an underestimate
of the H column density toward $\delta$ Sco.  Apart from this
sightline, the spread in N/H is comparable to those found
for Kr/H, O/H (Meyer et al.\ 1998), and C/H (\cite{car96,sof97})
with GHRS\@.  In any case, the $\delta$ Sco discrepancy is small
enough that omitting this sightline from the sample would only
slightly reduce the weighted mean N abundance from 
10$^6$ N/H $=$ 75 $\pm$ 4 to 73 $\pm$ 5.  The bottom line is
that the GHRS measurements yield an interstellar nitrogen
abundance that is about 80\% of the solar value with no
statistically significant variations from sightline to sightline.

As discussed by Meyer et al.\ (1998), a subsolar abundance
pattern in the local ISM today implies that something unusual
happened to either the Sun or the local ISM in the context of
standard Galactic chemical evolution models which predict that
the ISM metallicity should slowly increase over time.  The fact
that the GHRS interstellar abundances of C, N, O, and Kr vary
little from sightline to sightline makes it difficult to understand
this anomaly simply in terms of a typical ISM abundance fluctuation.
Possible explanations include the early enrichment of the solar
system by a local supernova (\cite{ree78,lee79,oli82}), a recent
infall of metal-poor gas in the local Milky Way
(\cite{com94,mey94,roy95}), or an outward diffusion of the Sun
from a birthplace at a smaller galactocentric distance
(\cite{wie96}).  A key prediction of the infall model is
that the mixture of metal-poor gas with the local ISM would lower
the abundances of all of the heavy elements below their solar
values by a similar amount.  The supernova enrichment hypothesis,
on the other hand, would create uneven elemental overabundances
in the Sun relative to the ISM that would reflect the
nucleosynthetic yields of one or more supernova events.
For example, the relative yield of O to N in Type II
supernovae (\cite{oli82}) is appreciably greater than their
relative present-day interstellar abundances.

If the solar N abundance and the N I $\lambda\lambda$1160,1161
oscillator strengths are accurate, the GHRS observations imply
that nitrogen is somewhat more abundant in the ISM than the 2/3
solar values measured for oxygen and krypton
(Meyer et al.\ 1998, Cardelli \& Meyer 1997).  This N enhancement
is illustrated in Figure 2 in terms of the N/Kr abundance ratio.
Although it should be small, the presence of any N in grains can
only serve to push this ratio (or N/O) further from the
equal deficit (with respect to the solar abundances) fiducial.
Thus, it would appear that nitrogen presents a problem
for the constant subsolar ISM abundance pattern predicted by the
infall model.  Furthermore, a higher value of N/O in the present-day
ISM than in the Sun is what one might expect if the protosolar
nebula was enriched by a local Type II supernova.  However, these
conclusions are not yet definitive because the
solar abundances and the N I $\lambda\lambda$1160,1161 $f$-values
are still uncertain enough that neither a subsolar ISM N abundance
similar to that of O and Kr or a solar ISM abundance can
be ruled out.  Indeed, the quality of the
GHRS data is now high enough that the limitations in comparing the
interstellar C, N, O, and Kr abundances no longer lie in the
measurements themselves but in the accuracy of the weak line
oscillator strengths and the solar abundances.

Defining an accurate set of ISM elemental abundances is also
important in determining the composition of interstellar dust grains.
Based on the B-star CNO abundances and the GHRS data on O and Kr
in the ISM, a general consensus has been developing that a subsolar
B-star standard may be the most appropriate for this work
(Sofia et al.\ 1994, Savage \& Sembach 1996, Snow \& Witt 1996).  
However, applying this standard to GHRS
measurements of the interstellar gas-phase
carbon abundance (\cite{car96,sof97}) yields a C dust fraction
(10$^6$ C/H $\approx$ 100) that is appreciably smaller than that
typically required (10$^6$ C/H $\approx$ 300) by models to
explain the total optical/UV dust opacity (\cite{mat89,sie92,kim94}).
\cite{mat96} has recently developed a model that reduces this
solid carbon requirement to 10$^6$ C/H $\approx$ 150 and other low-C
models may soon follow.  If N/O is indeed overabundant in the ISM
with respect to the Sun, the same could also be true of C/O and
thus somewhat relax the carbon constraints on these models.
Such a C/O overabundance would be expected in the scenario where
the early solar system is enriched by a nearby Type II supernova
(\cite{oli82}).  In any case, our GHRS observations of interstellar
nitrogen allow for the possibility that at least some elements do
not follow the same subsolar abundance pattern set for the ISM
by O and Kr.

\acknowledgments

This work was supported by STScI through a grant to Northwestern
University.

\clearpage

\figcaption[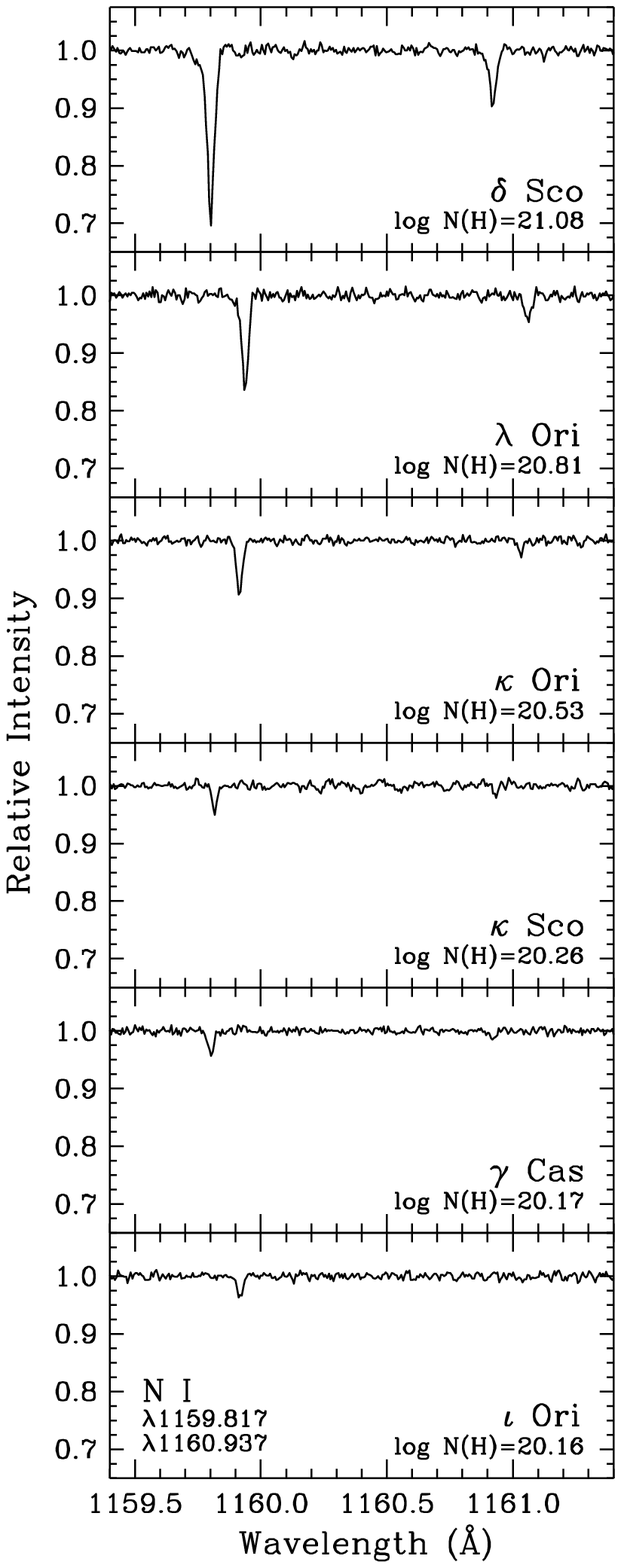]{{\it HST} GHRS echelle spectra of the interstellar
N I $\lambda\lambda$1159.817,1160.937 absorption doublet toward
$\delta$ Sco, $\lambda$ Ori, $\kappa$ Ori,
$\kappa$ Sco, $\gamma$ Cas, and $\iota$ Ori at a velocity
resolution of 3.5 km s$^{-1}$.  The normalized spectra are
displayed from top to bottom in order of decreasing total
hydrogen column density in the observed sightlines.  The measured
S/N ratios of these spectra are all in the 150 - 250 range.
The measured equivalent widths of the N I lines are listed in
Table 1.}

\figcaption[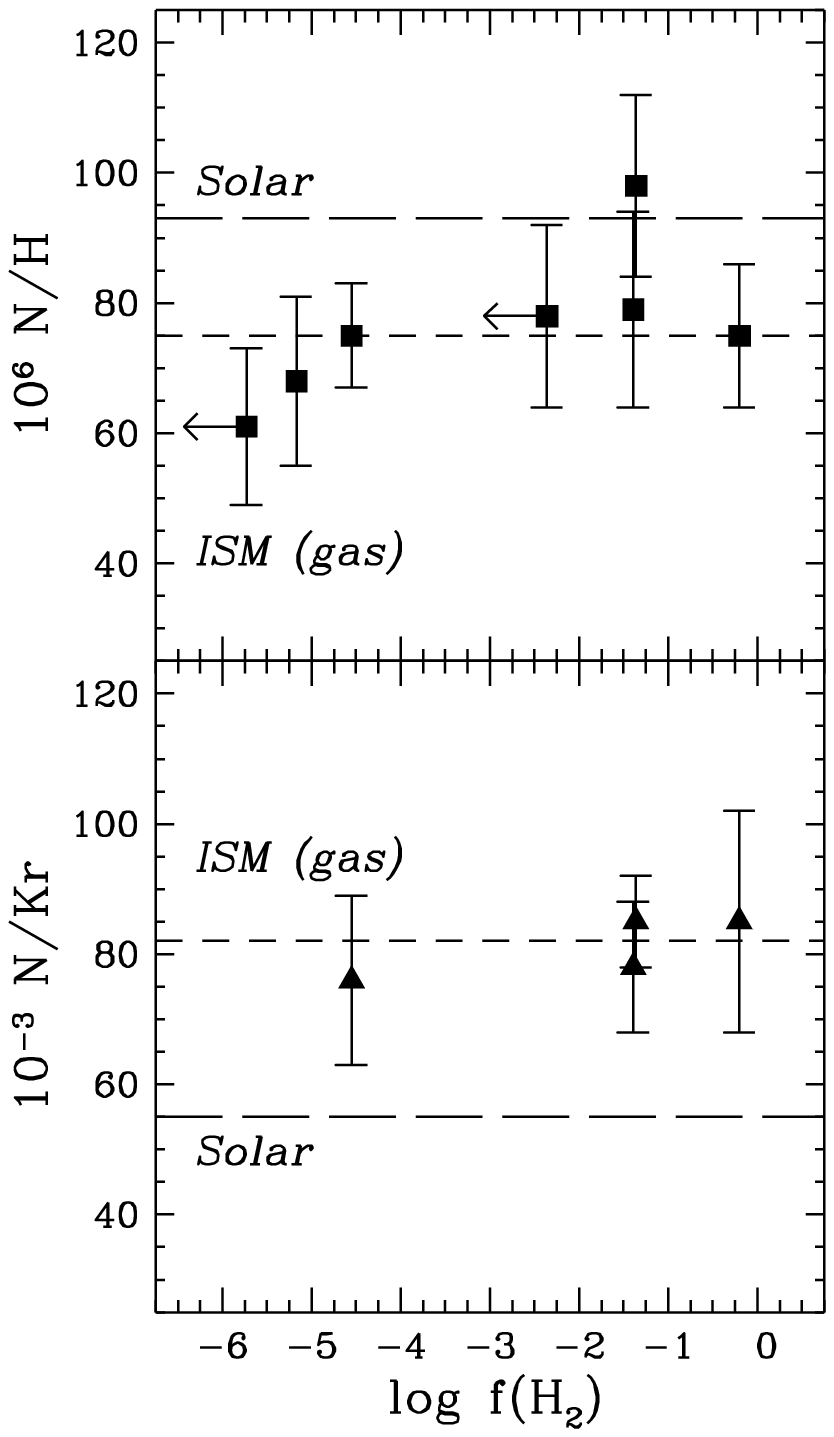]{Interstellar nitrogen abundances measured
with GHRS as a function of the logarithmic
fraction of hydrogen in molecular form,
$f$(H$_2$) $=$ 2$N$(H$_2$)/$N$(H),
in the observed sightlines.  In the top panel, the N abundances
are plotted in terms of 10$^6$ N/H as taken from Table 1.  The
short-dashed line among the data points represents the weighted
mean interstellar gas-phase N abundance (per 10$^6$ H atoms) of
10$^6$ N/H $=$ 75 $\pm$ 4.  This N abundance is about 80\% of the
Grevesse \& Noels (1993) solar value (10$^6$ N/H $=$ 93 $\pm$ 16)
represented by the long-dashed line.  In the bottom panel, the
N/Kr abundance ratio is plotted for the four sightlines in
common between this paper and the Kr study of Cardelli \& Meyer
(1997).  The short-dashed line among the data points represents
the weighted mean interstellar gas-phase N/Kr abundance ratio
of 10$^{-3}$ N/Kr $=$ 82 $\pm$ 5.  The solar value of N/Kr
(10$^{-3}$ N/Kr $=$ 55 $\pm$ 13) represented by the long-dashed
line incorporates the solar Kr abundance measured by Anders \&
Grevesse (1989).}

\clearpage

\begin{deluxetable}{lcccccc}
\footnotesize
\tablecaption{The GHRS Interstellar Nitrogen Abundances
\label{tbl2}}
\tablewidth{0pt}
\tablehead{
\colhead{Star} & \colhead{$N$(H)\tablenotemark{a}} &
\colhead{log $f$(H$_2$)\tablenotemark{b}} &
\colhead{$W_\lambda$(1160)\tablenotemark{c}} &
\colhead{$W_\lambda$(1161)\tablenotemark{c}} &
\colhead{$N$(N I)\tablenotemark{d}} &
\colhead{10$^6$ N/H\tablenotemark{e}}\\
\colhead{} & \colhead{(cm$^{-2}$)} &
\colhead{} & \colhead{(m\AA)} & \colhead{(m\AA)} &
\colhead{(cm$^{-2}$)} & \colhead{}}
\startdata
$\iota$ Ori & 1.5 (0.2) $\times$ 10$^{20}$ & $-$5.17
& \phn1.00 (0.15) & $<$ 0.30 & 9.87 (1.48) $\times$ 10$^{15}$
& 68 (13) \nl
$\gamma$ Cas & 1.5 (0.2) $\times$ 10$^{20}$ & $<-$2.36
& \phn1.15 (0.12) & 0.40 (0.12) & 1.15 (0.11) $\times$ 10$^{16}$
& 78 (14) \nl
$\kappa$ Sco & 1.8 (0.3) $\times$ 10$^{20}$ & $<-$5.73
& \phn1.10 (0.15) & 0.40 (0.15) & 1.11 (0.14) $\times$ 10$^{16}$
& 61 (12) \nl
$\kappa$ Ori & 3.4 (0.3) $\times$ 10$^{20}$ & $-$4.55
& \phn2.60 (0.15) & 0.65 (0.15) & 2.54 (0.14) $\times$ 10$^{16}$
& 75 (\phn8) \nl
$\lambda$ Ori & 6.5 (1.2) $\times$ 10$^{20}$ & $-$1.39
& \phn4.95 (0.20) & 1.45 (0.20) & 5.15 (0.30) $\times$ 10$^{16}$
& 79 (15) \nl
$\delta$ Sco & 1.2 (0.2) $\times$ 10$^{21}$ & $-$1.36
& 11.20 (0.30) & 3.30 (0.30) & 1.17 (0.06) $\times$ 10$^{17}$
& 98 (14) \nl
$\zeta$ Oph & 1.4 (0.1) $\times$ 10$^{21}$ & $-$0.20
& \phn7.56 (0.74) & 2.68 (0.99) & 1.05 (0.13) $\times$ 10$^{17}$
& 75 (11) \nl
\enddata
\tablenotetext{a}{$N$(H) $=$ 2$N$(H$_2$) $+$ $N$(H I) is the total
hydrogen column density ($\pm$1$\sigma$) in the observed sightlines.
These values reflect the H$_2$ column densities measured by
\cite{sav77} (\cite{jen86} in the case of $\kappa$ Sco) and the
weighted means of the \cite{boh78} (Jenkins et al.\ 1986 in the
case of $\kappa$ Sco) and \cite{dip94} $N$(H I) data.}
\tablenotetext{b}{$f$(H$_2$) $=$ 2$N$(H$_2$)/$N$(H) is the fractional
abundance of hydrogen nuclei in H$_2$ in the observed sightlines.}
\tablenotetext{c}{The measured equivalent widths ($\pm$1$\sigma$) of
the N I 1159.817 and 1160.937 \AA\ absorption lines.  The value
listed for $\lambda$1161 toward $\iota$ Ori is a 2$\sigma$ upper
limit.}
\tablenotetext{d}{The derived N I column densities ($\pm$1$\sigma$)
in the observed sightlines.  The $\zeta$ Oph value is
taken from the analysis of \cite{sav92}.  The
$\lambda$ Ori and $\delta$ Sco values are corrected for
a slight amount of saturation using respective Gaussian
$b$-values ($\pm$1$\sigma$) of
5.0$^{+\infty}_{-2.5}$ and 10.0$^{+\infty}_{-5.0}$
km s$^{-1}$.  The other sightlines are assumed to be optically thin
in the N I transitions.}
\tablenotetext{e}{The abundance of interstellar gas-phase nitrogen
($\pm$1$\sigma$) per 10$^6$ H atoms in the observed sightlines.  The
uncertainties reflect the propagated $N$(H) and $N$(N I) errors.}
\end{deluxetable}

\end{document}